\magnification \magstep 1
\hoffset -4truemm
\font\teneul=eufm10
\font\seveneul=eufm7
\font\fiveeul=eufm6
\newfam\eulfam
\textfont\eulfam=\teneul  \scriptfont\eulfam=\seveneul
  \scriptscriptfont\eulfam=\fiveeul
\def\eu{\fam\eulfam\teneul}
\font\tensanserif=cmss10

\def\ssf{\tensanserif}
 2
\parskip=7pt plus 1pt

\def\F{{\cal F}}

\def\o{{\bf 0}}
\def\p{{\bf p}}
\def\q{{\bf q}}
\def\L{{\cal L}}

\def\QED {\hfill\break 
     \line{\hfill{\vrule height 1.2ex width 1.2ex }\quad} 
      \vskip 0pt plus36pt} 
\parskip=14pt
\def\A{{\eu A}}
\def\B{{\eu B}}

\def\Cx{{\bf C}}

\def\G{{\eu G}}
\def\H{{\eu H}}
\def\K{{\eu K}}

\def\Nl{{\bf N}}

\def\Rl{{\bf R}}

\def\X{{\eu X}}

\def\Ir{{\bf Z}}

\def\idty{{\leavevmode{\rm 1\ifmmode\mkern -5.4mu\else
                                            \kern -.3em\fi I}}}

\overfullrule=0pt

\def\avg{\mathop{\rm Avg}\nolimits}
\def\bavg{\mathop{\hbox{\bf Avg}}\nolimits}
\def\mod{\,\mathop{\rm mod}\nolimits\,}
\def\nl{\hfill\break}
\parindent=20pt
\def\lessblank{\parskip=0pt}

{\nopagenumbers
\font\BF=cmbx10 scaled \magstep 3
\hrule height 0pt
\vskip 3\baselineskip

\centerline{\BF Multi-time correlations in}
\vskip \baselineskip
\centerline{\BF relaxing quantum dynamical systems}
\vskip 2\baselineskip
\centerline{Johan~Andries\footnote{$^1$}{Email: {\tt
             johan.andries@fys.kuleuven.ac.be}}, 
            Fabio~Benatti\footnote{$^2$}{Permanent address: Dept. Theor.
	    Phys. University of Trieste, Italy,} 
            \footnote{}{Email: {\tt
	    benatti@ts.infn.it}}$\!\!$, 
            Mieke~De~Cock\footnote{$^3$}{Onderzoeker FWO, Email: {\tt
	    mieke.decock@fys.kuleuven.ac.be}},
	    Mark~Fannes\footnote{$^4$}{Onderzoeksleider FWO, Email:  
            {\tt mark.fannes@fys.kuleuven.ac.be}}  
            } 
\vskip \baselineskip

\centerline{Instituut voor Theoretische Fysica} 
\centerline{Katholieke Universiteit Leuven} 
\centerline{Celestijnenlaan 200D} 
\centerline{B-3001 Heverlee, Belgium} 
\vskip 3\baselineskip plus 30pt

\centerline{\bf Abstract}
In this paper, we consider the long time asymptotics of multi-time
correlation functions for quantum dynamical systems that are
sufficiently random to relax to a ``reference state''. In particular,
the evolution of such systems must have a continuous spectrum. Special 
attention is paid to general dynamical clustering conditions and their
consequences for the structure of fluctuations of temporal averages. 
\vskip 2\baselineskip

\vfill\eject}

\pageno=1

\beginsection{1 Introduction}

One of the main problems in quantum chaos is to understand the
relaxation phenomena induced by the dynamics in systems with few
degrees of freedom. Typically, the system relaxes on an appropriate
time scale characteristic of the dynamics. Indeed, the spectrum of  the
evolution of chaotic quantum systems is usually discrete and
observation of the system for very long times will reveal this discrete
nature: the time correlation functions are quasi-periodic. The
separation between the points of the spectrum depends on a quantization
parameter such as $\hbar$ in the Chirikov kicked rotator~[1] or on
the dimension of an irreducible representation of SU(2) such as in the
kicked top~[2] or on that of an underlying Hilbert space as in the
finite-dimensional Cat maps~[3, 4]. When the quantization parameter
tends to an appropriate limit, one obtains a classical dynamical
system. We are not concerned with the statistical properties of
eigenvalues and eigenvectors of such models, but rather with dynamical
properties of expectations values. It is a typical feature that the
classical limit and the limit for large times of such expectation
values cannot be exchanged and so one can look for a joint limit
obtained by a suitable scaling of the time with respect to the
quantization parameter~[5]. The aim is to extract by such a procedure
true relaxation in rescaled time.

We don't address this scaling problem in this paper, but rather
concentrate on the large time behaviour of model systems with already
fully displayed relaxation as a consequence of basic dynamical
randomness. Different types of behaviour are possible and we introduce
a way of describing them by considering the asymptotic analysis of
multi-time  correlation functions~[6]. In particular it appears that
fluctuations around temporal averages are a useful tool to distinguish
between various degrees of randomness. For instance, the distribution
of fluctuations may be given by the usual Gaussian law, but also by the
semicircle law or by more exotic laws. Such distributions have
already been  obtained considering fluctuations of observables obeying
stochastic commutation  relations~[7]. In this paper, we show that
deterministic dynamics, i.e.\ without any stochastic input, can lead to
quite a variety of distributions for fluctuations. The appearance of
unusual statistics, such as the free statistics associated with the
semicircle law, is connected with chaotic features of the dynamics.
By chaotic or random quantum features, we shall mean different degrees of 
clustering in time, namely different strengths with which events largely
separated in time tend to become independent. 
Classically, one has the notion of mixing, whereas in quantum mechanics two 
notions are commonly considered, that of weak and strong clustering, the latter
one implying the former.
It turns out that the appearance of exotic statistics at the asymptotic level,
rater than the common, Gaussian one, associated with the notion of
classical independence and arising from the stronger clustering properties,
is related to a finer distinction of possibilities between weak and strong
clustering.
We anyway expect to observe the emergence of such statistics, not only in 
dynamical systems with fully displayed relaxation, but also for 
fluctuations in appropriately scaled finite systems.  

In Section~2, we review some notions of randomness for quantum
dynamical systems. Section~3 deals with the construction of the
asymptotics of multi-time correlation functions and provides hereby  a
useful setting for describing the law of large numbers for a
sufficiently ergodic dynamical system. Then, we consider in Section~4
how a central limit theorem for fluctuations can be obtained. In
Sections~5, we present the structures that arise when considering some
simple examples  of dynamical systems. 

In this paper, we shall model quantum dynamical systems by triples 
$(\A,\Theta,\phi)$ where
{\lessblank
\item{$\bullet$} 
 $\A$ is a unital C*-algebra of observables 
\item{$\bullet$} 
 $\Theta= \{\Theta_t\mid t\in\Rl\ \hbox{or } \Ir\}$ is a dynamical group of 
 $\ast$-automorphisms of $\A$ either in continuous or in discrete time and 
$X(t)$ 
 will denote the observable $X\in\A$ evolved up to time $t$: $X(t)= 
\Theta_t(X)$, 
 and  
\item{$\bullet$}
 the reference state $\phi$ is assumed to be invariant under $\Theta$ i.e.\ 
 $\phi\circ\Theta_t= \phi$. 

\noindent
General multi-time correlation functions are functions of the form }
$$
  {\bf t}\mapsto \phi\Bigl(X^{(1)}(t_{\nu(1)}) X^{(2)}(t_{\nu(2)}) 
  \cdots X^{(n)}(t_{\nu(n)})\Bigr)\ , 
\eqno(1)  
$$
where the $X^{(j)}(t_{\nu(j)})$ are operators at times $t_{\nu(j)}$ in
$\A$, ${\bf t}=\{t_1, t_2,\ldots\}\in \Ir^{\Nl_0}$ and $\nu$ maps
$\{1,2, \ldots, n\}$ into  $\Nl_0$. 

Usually, in quantum statistical mechanics, one considers time-ordered 
correlation functions. Since one expects observables largely separated
in time to commute, such an ordering would be no restriction. We are
interested, however, in a situation where a complex dynamics  generates
wildly fluctuating algebraic relations for observables largely 
separated in time. In such a case, commutation relations cannot be used
to  simplify correlation functions by grouping together observables at
equal large times and, in order to obtain information about algebraic relations
between observables largely separated in time, we have to consider general 
expressions as in~(1). In particular, it is
necessary  to include the possibility of repeating a  same time label
within a correlation function. Then, the natural algebraic  structure
to consider is that of  a countable free product  $\A_\infty$ of 
copies of $\A$~[8], while the asymptotics of  multi-time correlation
functions will allow us to  equip $\A_\infty$ with an {\ssf asymptotic
state} $\phi_\infty$. The probabilistic structure corresponding  to
such a state should reflect the essential features of  the
underlying dynamics.

We briefly remind the construction of $\A_\infty$ as the free product
$\star_{i\in\Nl_0} \A_i$ and we shall refer to $\A_\infty$ as the {\ssf 
asymptotic free algebra} associated with the dynamical system 
$(\A,\Theta,\phi)$. Each of the algebras $\A_i$ is a copy of the basic 
algebra $\A$ of observables and $\A_\infty$ is the universal 
C*-algebra generated by an identity 
element $\idty$ and by ``words'' $w= X^{(1)}_{\nu(1)} X^{(2)}_{\nu(2)} 
\cdots X^{(n)}_{\nu(n)}$ that consist of concatenations of ``letters'' 
$X^{(j)}\in \A$. The subscript $\nu(j)$ in $X^{(j)}_{\nu(j)}$ refers 
to which copy of $\A$ the letter $X^{(j)}$ belongs to and any two
consecutive subscripts are unequal. 
Concatenation, together with simplification rules, defines the product of words. 
More specifically, the rules for handling words are: for $X,\,Y\in\A$, 
$\lambda\in\Cx$, 
$j\in\Nl_0$ and $w,\,w'$ two generic words
$$\eqalignno{
  w \idty_j w'
  &\Rightarrow w w' &(2.{\rm a}) \cr
  w \bigl(X_j+ \lambda Y_j\bigr) w'
  &\Rightarrow w X_j w' + \lambda w Y_j w' &(2.{\rm b}) \cr
  w X_j Y_j w' 
  &\Rightarrow w (XY)_j w'\ . &(2.{\rm c}) 
}$$
Notice that the product $XY$ in~(2.c) is not concatenation, but
rather the usual operator product in the algebra $\A$. Moreover, the
adjoint $w^*$ of a word $w= X^{(1)}_{\nu(1)} X^{(2)}_{\nu(2)}  \cdots
X^{(n)}_{\nu(n)}$ equals $\bigl(X^{(n)\ast}\bigr)_{\nu(n)}
\bigl(X^{(n-1)\ast}\bigr)_{\nu(n-1)}  \cdots
\bigl(X^{(1)\ast}\bigr)_{\nu(1)}$.

\beginsection{2 Random behaviours in quantum systems}

Before considering the problem of endowing  $\A_\infty$ with an
asymptotic state, we present a hierarchy of ergodic properties typical
for infinite quantum systems~[9, 10]. We shall formulate them for the
case of discrete time dynamical systems $(\A,\Theta,\phi)$.

Actually, our construction will be essentially based on the use
of time averages of correlation functions of the form
$$
  \overline{\phi(X\, Y(t)\, Z)}^{\, \rm av}:=
  \lim_{T\to \infty} {1\over T} \sum_{s=0}^{T-1} \phi(X\, Y(s)\, Z)\ , 
\eqno(3)
$$ 
where $X$, $Y$ and $Z$ are arbitrary observables in $\A$.

If the dynamics of a system is sufficiently regular, observations turn
out to be quite correlated, even when largely separated in time. We
shall then associate various degrees of randomness with the strength of
decorrelation properties of the dynamics, if any. The lowest degree of
randomness is {\ssf clustering in the mean}
$$
  \overline{\phi(X\, Y(t)\, Z)}^{\, \rm av}= \phi(X\, Z)\, \phi(Y), 
  \qquad X,\,Y,\,Z\in\A \ , 
\eqno(4)
$$
next in the list comes {\ssf weak clustering}
$$
  \lim_{t\to\infty} \phi(X\, Y(t)\, Z)= \phi(X\, Z)\, \phi(Y), \qquad X,\, 
  Y,\, Z\in\A \ , 
\eqno(5)
$$
and we close the list with {\ssf strong clustering}
$$\lim_{t\to\infty} \phi(X\, Y(t)\, Z\, S(t)\, T)= \phi(X\, Z\, T)\, \phi(Y\, 
S),
  \qquad X,\, Y,\, Z,\, S,\,T\in\A
\eqno(6.{\rm a})
$$
and {\ssf hyper-clustering}
$$
  \lim_{\inf |t_i-t_j|\to\infty}\ 
  \phi\Bigl(X^{(1)}(t_{\nu(1)}) 
  X^{(2)}(t_{\nu(2)}) 
  \cdots X^{(n)}(t_{\nu(n)})\Bigr)= \prod_j 
\phi\Bigl(\overrightarrow{\prod}_{\kappa\in 
  \nu^{-1}(j)} X^{(\kappa)}\Bigr)\ .
\eqno(6.{\rm b})
$$ 
In formula~(6.b), the limit is taken in such a way that all the times
$t_{\nu(j)}$ and the differences between those with different indices go to 
infinity. Notice that, as
in~(1), repeated times are allowed in~(6.b) and the arrow over the product at 
the right-hand side indicates that the order among operators at equal times 
has to be preserved since they do not commute in general. 
Among the previous properties, one can check that the following relations hold
$$
  (6.{\rm b})\ \Leftrightarrow\  (6.{\rm a})\ \Rightarrow\
  (5)\         \Rightarrow\ 
  (4)\ .
\eqno(7)
$$			  
While the implications from left to right are evident, in order to
deduce the equivalence in the first place, we use the notion of {\ssf
asymptotic Abelianess} in time. Indeed, the different kinds of
clustering~(4), (5) and (6.a), imply the following degrees of
asymptotic Abelianess: {\ssf asymptotic Abelianess in the mean}
$$
  (4)\Longrightarrow
  \overline{\phi(S\, [X, Y(t)]\,Z)}^{\, \rm av}= 0 \qquad\forall
  S,\,X,\,Y,\,Z\in\A\ ,
\eqno(8.{\rm a})
$$ 	  
{\ssf weak asymptotic Abelianess}
$$
  (5)\Longrightarrow
  \lim_{t\to\infty} \phi(S\, [X, Y(t)]\,Z)=0 \qquad\forall S,\,X,\,Y,\,Z\in\A\, 
\eqno(8.{\rm b})
$$
and {\ssf strong asymptotic Abelianess}
$$
  (6.{\rm a})\Longrightarrow
  \lim_{t\to\infty} \phi(S^*\, [X,Y(t)]^*[X,Y(t)] S)=0 \qquad\forall\,
  S,\,X,\,Y\in\A\ .
\eqno(8.{\rm c})
$$
Weak clustering plus strong asymptotic Abelianess imply strong clustering. 
Indeed
$$\eqalign{
  \lim_{t\to\infty} \phi(X\, Y(t)\, Z\, S(t)\, T)
  &= \lim_{t\to\infty} \phi(X\, Y(t)\, S(t)\, Z\,T)+ \lim_{t\to\infty} 
  \phi(X\, Y(t)\, [Z,S(t)]\, T) \cr
  &= \phi(X\,Z\, T)\, \phi(Y\,S)\qquad\forall\, S,\, T,\, X,\, Y,\, Z\in\A\ . 
}$$
The second limit tends to zero because of 
strong asymptotic Abelianess and the Cauchy-Schwartz inequality
$$\eqalign{
  |\phi(X\,Y(t) [Z, S(t)]\, T)|^2
  &\le \phi(X\, Y(t)\, Y(t)^* X^*)\, \phi(T^*[Z,S(t)]^* [Z,S(t)]\, T) \cr
  &\le \|X\|^2\, \|Y\|^2\, \phi(T^*[Z,S(t)]^*[Z,S(t)]\, T)\ ,
}$$
while the first one gives the result because of weak-clustering.
By a similar argument, (6.a) implies~(6.b) and therefore the 
equivalence in~(7) holds. We just sketch here the main idea of the argument
by considering the case of three equal times. In a first step, for fixed
$X,\,Y,\,Q,\,S,\,T,\,U$ and $Z$ in $\A$, write
$$\eqalignno{
  \phi(X\,Y(t)\,S\,Z(t)\,T\,U(t)\,Q)
  &= \phi(X\,S\,Y(t)\,Z(t)\,T\,U(t)\,Q)+ \cr 
  &\hskip 2cm \phi(X\,[Y(t),S]\,Z(t)\,T\,U(t)\,Q)\ .
  &(9)
}$$
By the Cauchy-Schwarz inequality and $\phi(abb^*a^*)\le \|b\|^2\,
\phi(aa^*)$ we obtain
$$
  |\phi(X\,[Y(t),S]\, Z(t)\, T\, U(t)\, Q)|^2
  \le \|Z\|^2\, \|T\|^2\, \|U\|^2 \|Q\|^2\, \phi(X\,[Y(t),S][Y(t),S]^*
  X^*)\ .
$$
Assuming strong asymptotic Abelianess~(8.c), the second term
in~(9) vanishes when $t$ tends to infinity. Repeating the same
procedure on the first term, we collect together $(Y Z U)(t)$ and
the conclusion follows from weak-clustering.
Indeed, the condition that $\inf|t_i-t_j|\to\infty$ means that all different
times are so largely separated that the regrouped correlation functions
cluster asymptotically with respect to any of them.
		  
\beginsection{3 The asymptotic multi-time correlations}

Consider, as in~(1), a multi-time correlation function 
$$
  {\bf t}\mapsto \phi\Bigl(X^{(1)}(t_{\nu(1)}) X^{(2)}(t_{\nu(2)}) 
  \cdots X^{(n)}(t_{\nu(n)})\Bigr)\ . 
\eqno(10)
$$
We may restrict the map $\nu: \{1,2, \ldots, n\}\to \Nl_0$  in such a
way that if a $\nu(j)$ appears somewhere in the operator
product in~(10), then all smaller natural numbers must have already
appeared at least once to the left of $\nu(j)$. More formally
$$
  \hbox{if } 1\le\ell<\nu(j),\ \hbox{then there exists } 1\le i<j\ 
  \hbox{such that } \nu(i)=\ell\ .
\eqno(11)
$$
As a consequence of this prescription, $\nu(\{1,2, \ldots, n\})= \{1,2,
\ldots,k\}$ with $k\le n$ due to possible repetitions of a label. 
Notice that this is a convenient way of rewriting the multi-time
correlation functions that were introduced in~(1): e.g.\ ${\bf
t}\mapsto \phi(X(t_3) Y(t_2))$ is not excluded by the rule of above for
$\nu$ since it equals ${\bf t}\mapsto \phi\bigl(\idty(t_1) \idty(t_2))
X(t_3) Y(t_2)\bigr)$. The reason for making this specific choice in
writing correlation functions is to have all labels $1,2,\ldots, n$
appear, in increasing order and possibly with repetitions, in a
correlation function if the label $n$ appears. This will become useful
later on when we consider time averages. The multi-time correlation
functions~(10) form a self-adjoint linear space containing the constant
function.  

We shall describe the asymptotics of multi-time correlation functions in
terms of a state $\phi_\infty$ on the asymptotic free algebra 
$\A_\infty$. $\phi_\infty$ is obtained by averaging multi-time correlation
functions.
We now generalize the procedure of averaging considered in~(3) for a
single-time correlation function and briefly recall the notion of {\ssf
invariant mean}~[11]. A mean $\avg$ on a set $X$ is a {\ssf normalized}
and {\ssf positive} linear functional on a  self-adjoint linear space
$\X$ of bounded complex-valued functions on $X$  containing the
constant function and closed for the $\|\ \|_\infty$-norm. If $\G$ is a
group of transformations of $X$, leaving $\X$ globally invariant,
then   $\avg$ is called {\ssf invariant} if $\avg(f)= \avg (f\circ
\gamma)$.

As we consider discrete time dynamical systems $(\A,\Theta,\phi)$,
we take $\Ir^{\Nl_0}$ for the space $X$. 
The group $\G$ will consist of multi-time translations on
$X$ given by
$$
  \gamma_{\bf s}: X\to X: {\bf t}\mapsto {\bf t-s}\ , \qquad {\bf
  s}=\{s_1,s_2, \ldots\}\in\Ir^{\Nl_0}\ . 
$$   
The function space $\X$ on which we shall consider means is the closure
of the linear space of multi-time correlation functions. $\X$ is
invariant under the group action $f\mapsto f\circ \gamma_{\bf s}$ since
$\Theta$ is an automorphism of $\A$.

In general, we want to explore the possibility of defining a positive,
normalized linear functional (a state) $\phi_\infty$ on the linear span
of the words in the asymptotic free algebra, via some mean
``$\bavg$ '' defined on the space of multi-time correlation functions.

We proceed as follows: first we rewrite a general word $w=
X^{(1)}_{\mu(1)} X^{(2)}_{\mu(2)} \cdots X^{(n)}_{\mu(n)}$ without
restrictions on the map $\mu$ except that consecutive subindices
$\mu(j)$ and $\mu(j+1)$ are different. By inserting an appropriate
number of identities we can write in a unique way $w= Y^{(1)}_{\nu(1)}
Y^{(2)}_{\nu(2)} \cdots Y^{(r)}_{\nu(r)}$, where the $Y$ are either
identities or are $X$'s and where $\nu$ satisfies~(11). E.g.\ $w= X_2$
is rewritten as $w=\idty_1 X_2$ and $w= X_2 Y_1 Z_2 W_4$ becomes $w=
\idty_1 X_2 Y_1 Z_2 \idty_3 W_4$.  Given the space of multi-time
correlation functions of a dynamical system $(\A,~\Theta,\phi)$, we
shall consider, when existing, the linear  functionals
$$
  \phi_\infty\Bigl(X^{(1)}_{\nu(1)} X^{(2)}_{\nu(2)} \cdots
  X^{(n)}_{\nu(n)}\Bigr):= \bavg\Bigl({\bf t}\mapsto
  \phi\Bigl(X^{(1)}(t_{\nu(1)}) X^{(2)}(t_{\nu(2)}) \cdots
  X^{(n)}(t_{\nu(n)})\Bigr)\Bigr)\ ,  
\eqno(12)
$$
where $\nu$ maps $\{1,2,\ldots,n\}$ into $\{1,2,\ldots,s\}$.
Furthermore, we restrict ourselves to those means $\bavg$ on $\X$ that
satisfy the following {\ssf strong compatibility} condition:

{\narrower\noindent
 if  $(t_1, t_2, \ldots, t_n)\mapsto f(t_1, t_2, \ldots, t_n)$ is a
 uniform limit of multi-time correlation functions, and if  $f$ depends
 only on a subset $\{t_{i(1)}, t_{i(2)}, \ldots, t_{i(k)}\}$ of
 variables where $i: \{1,2,\ldots, k\}\to \{1,2,\ldots, n\}$ is an
 order preserving injection, that is $f(t_1, t_2, \ldots, t_n)=
 g(t_{i(1)}, t_{i(2)}, \ldots, t_{i(k)})$, then $\bavg(f)= \bavg(g)$.
 \par
}
\vskip-\baselineskip
\line{\hfill (13)}
\smallskip

\noindent
{\bf Proposition~1.}
{\it 
 If $\bavg$ exists as an invariant mean on the space of multi-time
 correlation functions and satisfies the strong compatibility
 condition~(13), then the functional $\phi_\infty$ defined
 in~(12) extends to a state on $\A_\infty$. Furthermore, 
 $$
   \phi_\infty\circ {\bf \Theta_s}= \phi_\infty 
   \qquad\hbox{and}\qquad
   \phi_\infty\circ \alpha_\theta= \phi_\infty \ , 
 $$
 where $\bf \Theta_s$ satisfies ${\bf \Theta_s}(X_j):=
 \bigl(\Theta_{s_j}(X)\bigr)_j$  for $j\in\Nl_0$ and $X\in\A$, whereas
 $\theta$ is any order preserving injective transformation of $\Nl_0$
 and $\alpha_\theta$ is the $\ast$-homomorphism of $\A_\infty$
 determined by $\alpha_\theta(X_j):= X_{\theta(j)}$.
}
  
\noindent
{\bf Proof:} 
 Normalization and linearity are a consequence of~(2.a)
 and~(2.b). In order to prove positivity we have to check that for
 any finite linear combination $W= \sum_w \lambda_w\, w$ of words
 one has $\phi_\infty(W^*W)\ge 0$. {From} definition~(12) it turns out that
 $\phi_\infty(W^*W)$
 is the multiple average of  the expectation of a positive operator in the
 state $\phi$, hence positive. Multi-time invariance follows from the
 invariance of the mean. Again, instead of producing a formal proof of the
 invariance of $\phi_\infty$ under $\alpha_\theta$, we present a simple
 example which clarifies the essential mechanism. We show that
 $\phi_\infty(X_2 Y_1 Z_2)= \phi_\infty(X_3 Y_1 Z_3)$, in which case
 $\theta(1)=1$ and $\theta(2)=3$. Relabelling dummy variables and
 using condition~(13), we obtain
 $$\eqalignno{
   \phi_\infty(X_3 Y_1 Z_3)
   &= \bavg\Bigl((t_1, t_2, t_3, \ldots)\mapsto \phi\bigl( \idty(t_1)
   \idty(t_2) X(t_3) Y(t_1) Z(t_3)\bigr) \Bigr) \cr
   &= \bavg\Bigl((t_1, t_2, t_3, \ldots)\mapsto \phi\bigl( \idty(t_1)
   X(t_2) Y(t_1) Z(t_2)\bigr) \Bigr) \cr
   &= \phi_\infty(X_2 Y_1 Z_2)\ .
   &\hbox{$\vrule height 1.2ex width 1.2ex $}\quad
 }$$

In the following, we shall fix a definite averaging procedure and construct
asymptotic states $\phi_\infty$ averaging over the different times in
successive order
$$\eqalignno{
  &\bavg\Bigl({\bf t}\mapsto f(t_1,t_2, \ldots, t_n)\Bigr):= \cr
  &\qquad \avg\Bigl(t_n\mapsto \cdots \avg\Bigl(t_2\mapsto
  \avg\Bigl(t_1\mapsto f(t_1, t_2, \ldots,
  t_n)\Bigr)\Bigr)\cdots\Bigr)\ .
  &(14)
}$$  
More general procedures are possible by considering coupled limits:
e.g.\ one could try to construct a mean by averaging functions
$(t_1, t_2, \ldots, t_n)\mapsto f(t_1,t_2, \ldots, t_n)$ of
$n$ arguments over $n$-dimensional cubes
$$
  \bavg\Bigl((t_1, t_2, \ldots, t_n)\mapsto f(t_1,t_2, \ldots,
  t_n)\Bigr):=  \lim_{L\to\infty} {1\over L^n} \sum_{t_1=1}^L\cdots
  \sum_{t_n=1}^L  f(t_1, \ldots, t_n) \ .
$$  
Such a mean is also strongly compatible. Notice, however, that, 
choosing different single-time averages $\avg$ in~(14), we can
in general violate the strong compatibility condition~(13). 
Here follows an example with two times and two different single-time 
averages
$$\eqalign{
  \avg_2\Bigl(t_2\mapsto \avg_1\Bigl(t_1\mapsto f(t_2)\Bigr)\Bigr)
  &= \avg_2\Bigl(t\mapsto f(t)\Bigr) \cr
  &\ne \avg_1\Bigl(t\mapsto f(t)\Bigr) \cr
  &= \avg_2\Bigl(t_2\mapsto \avg_1\Bigl(t_1\mapsto f(t_1)\Bigr)\Bigr)
  \ . 
}$$  

Quite to the other extreme, there are examples of dynamical systems
and of averages of their multi-time correlation functions that are not 
only strongly compatible but even {\ssf permutation invariant} in the 
sense that
$$
  \bavg\Bigl( \alpha_\pi(f)\Bigr)= \bavg(f)
\eqno(15)
$$ 
for any local permutation $\pi$ of the natural numbers, that is for any
bijection  $\pi$ from $\Nl_0$ into $\Nl_0$ that leaves all elements
invariant except for a finite number of them. $\alpha_\pi$ acts on a
correlation function ${\bf t}\mapsto f(t_1,t_2, \ldots)$ as
$\alpha_\pi(f)(t_1, t_2, \ldots)= f(t_{\pi(1)}, t_{\pi(2)}, \ldots)$.
As a consequence
of~(15),  the asymptotic states $\phi_\infty$ defined by permutation
invariant multi-time averages will also be permutation invariant on the
asymptotic algebra $\A_\infty$, namely
$$
  \phi_\infty\Bigl( X^{(1)}_{\nu(1)} X^{(2)}_{\nu(2)} \cdots 
  X^{(n)}_{\nu(n)}\Bigr)= 
  \phi_\infty \Bigl(X^{(1)}_{\pi\circ\nu(1)} X^{(2)}_{\pi\circ\nu(2)} \cdots
  X^{(n)}_{\pi\circ\nu(n)}\Bigr)\qquad\forall\, X^{(j)}\in\A\ .
\eqno(16)
$$ 

The asymptotic states $\phi_\infty$ are always permutation invariant when the 
basic dynamical system $(\A,\Theta,\phi)$ is strongly clustering.
In order to prove this, we introduce a useful technical result.

\noindent
{\bf Lemma~1.}
{\it
  For $d,k\in \Nl$ define
  $$\Delta_d^k (t_1,\ldots,t_k) = \cases{
   0 & if $|t_i - t_j|\leq d$ for some $1\leq i \not= j \leq k$ \cr
   1 & else}\ . 
  $$
  Then, if the multiple average~(14) of a uniformly bounded function 
  $f:\Ir^k\to \Cx$ exists, we have 
  $$\eqalign{
    &\avg\Bigl( t_k\mapsto \cdots \avg\Bigl( t_2\mapsto
    \avg\Bigl( t_1\mapsto f(t_1, t_2, \ldots, t_k)\Bigr)\Bigr)\cdots
    \Bigr)= \cr
    &\avg\Bigl( t_k\mapsto \cdots \avg\Bigl( t_2\mapsto
    \avg\Bigl( t_1\mapsto \Delta_d^k(t_1,\ldots,t_k) f(t_1, t_2,
    \ldots, t_k)\Bigr)\Bigr)\cdots \Bigr) \ .
  }$$ 
}

\noindent
{\bf Proof:} 
 For fixed $d$, choose $t_1,\ldots,t_{i-1},t_{i+1}, \ldots, t_k \in
 \Ir$ in such a way that the function
 $\Delta_d^{k-1}(t_1,\ldots,t_{i-1},t_{i+1},\ldots,t_k)= 1$. Then,
 $\Delta_d^k(t_1,\ldots,t_i,\ldots,t_k)= 0$ for only a finite number of
 values of $t_i \in \Ir$ which implies that   
 $$\eqalign{
   \avg\Bigl(t_i\mapsto \Delta_d^k & (t_1,\ldots,t_i,\ldots,t_k)\,
        g(t_1,\ldots,t_i,\ldots, t_k)\Bigr) \cr
    & = \Delta_d^{k-1} (t_1,\ldots,t_{i-1},t_{i+1},
   \ldots,t_k)\, \avg\Big(t_i\mapsto g(t_1,\ldots,t_i,\ldots,t_k)\Bigr)
 }$$
 for a uniformly bounded $g:\Ir^k\to\Cx$. By successively applying this
 observation to
 $$
   \avg\Bigl( t_n\mapsto \cdots \avg\Bigl( t_2\mapsto \avg\Bigl(
    t_1\mapsto \Delta_d^n(t_1,\ldots,t_n) f(t_1, t_2, \ldots,
    t_n)\Bigr)\Bigr)\cdots \Bigr)\ ,
 $$ 
 the lemma follows. Whereas in the statement of the lemma a definite
 order of the time averages has been specified, its proof is independent
 of it. 
\QED

\noindent
{\bf Proposition~2. }
{\it 
 Let $(\A,\Theta,\phi)$ be strongly clustering as in~(6.a), 
 equivalently~(6.b). Then, if the multiple
 average in~(14) exists, it defines a permutation invariant asymptotic state
 $\phi_\infty$ on $\A_\infty$.
}

\noindent
{\bf Proof:}
 Let us consider a multi-time correlation function
 $$
   \phi \Bigl(X^{(1)}(t_{\nu(1)})  X^{(2)}(t_{\nu(2)})  
    \cdots X^{(n)}(t_{\nu(n)})\Bigr)\ ,
 $$
 where, because of strong compatibility, we can assume that
 $\nu:\{1,2,\ldots,n\}\to \{1,2,\ldots,k\}$ with $k\le n$. Thus, we
 collect all $i\in\{1,2,\ldots,n\}$ such that  $\nu(i)=p$ into subsets
 $I_p$. Because of strong-clustering, and thus of
 hyper-clustering~(6.b),  for any $\epsilon>0$,  we can choose $d>0$
 such that $|t_i-t_j|\ge d$ for all $t_i\neq t_j$, with
 $i,j\in\{1,2,\ldots,k\}$, implies 
 $$
   \Bigl| \phi\Bigl(X^{(1)}(t_{\nu(1)}) X^{(2)}(t_{\nu(2)}) 
   \cdots X^{(n)}(t_{\nu(n)})\Bigr)-\prod_{p=1}^k \phi\Bigl(
   \overrightarrow{\prod}_{i\in I_p} X^{(i)}\Bigr)
   \Bigr|\le\epsilon\ .
 $$
 Thus, when evaluating  
 $\phi_\infty \Bigl(X^{(1)}_{\nu(1)}  X^{(2)}_{\nu(2)} 
  \cdots X^{(n)}_{\nu(n)}\Bigr)$
 via the prescription~(14), we can use Lemma~1 and   
 a function $\Delta_d^k(t_1,t_2,\ldots,t_k)$ to estimate
 $$
   \Bigl| \phi_\infty\Bigl(X^{(1)}_{\nu(1)} X^{(2)}_{\nu(2)} 
   \cdots X^{(n)}_{\nu(n)}\Bigr)-\prod_{p=1}^k \phi\Bigl(
   \overrightarrow{\prod}_{i\in I_p} X^{(i)}\Bigr)
   \Bigr|\le\epsilon\ .
 $$
 Therefore, in the case of strong clustering, in whichever order the
 single time averages are performed,  the multiple average~(14) agrees 
 with the time limit~(6.b) of multi-time
 correlation functions. The ensuing asymptotic state $\phi_\infty$ is
 thus permutation invariant.  Moreover, from time invariance of $\phi$,
 we have that  $\phi_\infty(X_j)= \phi(X)$ for any $j\in\Nl_0$ and
 $X\in\A$. Therefore, with $i\in I_p$ implying $\nu(i)=p$,
 $$
   \phi_\infty \Bigl(X^{(1)}_{\nu(1)}  X^{(2)}_{\nu(2)}  \cdots
   X^{(n)}_{\nu(n)}\Bigr)= \prod_{p=1}^k \phi_\infty\Bigl( 
   \overrightarrow{\prod}_{i\in I_p} X^{(i)}\Big)\ .
 \eqno(17)
 $$ 
 
\QED

When the basic dynamics is strongly clustering, the structure of the
expectations on $\A_\infty$ calculated with respect to the asymptotic
state $\phi_\infty$ corresponds to the usual notion of  {\ssf
commutative independence} of random variables.  Indeed,  $\phi_\infty$
vanishes on the two-sided ideal of $\A_\infty$ generated by commutators
of letters sitting in different copies of $\A$ in $\A_\infty$, i.e.\ by
$\{[X_j,Y_k]\mid j\ne k,\, X,\,Y\in\A\}$. We may therefore think of 
$\phi_\infty$ as an infinite product state on the minimal tensor
product of a countable number of copies of $\A$. 

A very different notion of independence, called {\ssf freeness} or
{\ssf free independence}, has been recently introduced in the realm
of non-commutative probability~[8]. This notion corresponds to the
following structure for the correlation functions of a state $\psi$ on
a free product $\star_j \B_j$ of C*-algebras $\B_j$
$$
  \psi\Bigl(X^{(1)}_{j_1} X^{(2)}_{j_2} \cdots X^{(n)}_{j_n}\Bigr)=0
$$ 
whenever $\psi\Bigl(X^{(k)}_{j_k}\Bigr)=0$ and $j_k\ne j_{k+1}$ for
all $k$. Accordingly, one refers to $\psi$ as to the free product of the
states $\psi_j$, where $\psi_j$ is the restriction of $\psi$ to $\B_j$.  

Freeness is totally incompatible with statistical independence as
in~(17). In fact, if $\phi_\infty$ satisfied both freeness and
commutative independence, then we would have for any centred
observable $X$, i.e.\ $\phi(X)=0$, that
$$
  0= \phi_\infty(X^*_0X^*_1X_0\,X_1)= \phi(X^*X)^2\ .
$$

\beginsection{4 Temporal fluctuations}

The two cases of commutative and free independence, presented in the
previous section, are somehow extreme. Many other possibilities for the
structure of $\phi_\infty$ may arise. In general, we cannot hope to
obtain a comprehensive description of these structures.  A more
manageable framework is provided by ``fluctuations''. They are natural
objects to consider as we may think of the state $\phi_\infty$ as
determining the Law of Large Numbers for the dynamical system
$(\A,\Theta,\phi)$.
Fluctuations are thus on the level of the Central
Limit Theorem.  

\noindent
{\bf Definition. }
{\it 
  Let $N\in \Nl_0$ and $X\in\A$. A local fluctuation $F_N(X)$ is the 
  following element of $\A_\infty$
  $$
    F_N(X):= {1\over \sqrt N} \sum_{i=1}^N \Bigl(X_i- \phi(X)\idty
    \Bigr)\ .
  $$  
}

It is our aim to discuss the limiting behaviour of $F_N(X)$ when $N$ tends
to infinity, namely to study limits of correlations such as
$$
  \lim_{N\to\infty} \phi_\infty\Bigl(F_N(X^{(1)}) F_N(X^{(2)}) 
  \cdots F_N(X^{(r)})\Bigr)\ , 
\eqno(18)
$$
and to establish an algebraic central limit theorem. 

In a stronger sense, one may try to reconstruct possible algebraic relations
of global fluctuations $F(X)$, if any, by means of the functional $\Phi$
on global fluctuations defined by
$$
  \Phi\Bigl(F(X^{(1)}) F(X^{(2)}) \cdots
  F(X^{(r)})\Bigr):= \lim_{N\to\infty} \phi_\infty\Bigl(F_N(X^{(1)}) 
F_N(X^{(2)}) 
  \cdots F_N(X^{(r)})\Bigr)\ . 
\eqno(19)
$$
Notice that, in the case of commutative, respectively free
independence, the linear functional $\Phi$ on the r.h.s.\ of~(19) is
in fact  a well-defined state on the algebra of fluctuations~[12, 7] 
such that they become Gaussianly, respectively
semicircularly, distributed, non-commutative, random variables. 
 
We shall restrict our considerations to averages of
multi-time correlation functions of the form~(14) and study
in some generality the limit~(18) by adapting an argument in~[13].
We show that a clustering condition stronger than weak clustering~(5), but 
weaker 
than strong clustering~(6.a), is sufficient to ensure that only
moments of even order contribute to the limit joint distribution of
fluctuations.

\noindent
{\bf Proposition~3.}
{\it 
 Let us assume that the quantum dynamical system  $(\A,\Theta,\phi)$
 satisfies the following cluster condition
 $$
   \lim_{\inf |t_i-t_j|\to\infty} \phi\Bigl(Z^{(1)}(t_{\nu(1)}) \cdots
   Z^{(j)}(t_{\nu(j)}) Y Z^{(j+1)}(t_{\nu(j+1)})\cdots
   Z^{(n)}(t_{\nu(n)})\Bigr)= 0\ ,
 \eqno(20)
 $$
 whenever $Y$ is centred and $\nu$ maps $\{1,2,\ldots,n\}$ into
 $\Nl_0$, where, as in~(6.b), $\lim_{\inf |t_i-t_j|\to\infty}$
 means that all times and differences of different times go 
 to infinity.
 If the multiple average~(14) exists, the state $\phi_\infty$ it
 defines on the asymptotic algebra $\A_\infty$ is such that, with
 $X^{(1)}, \ldots, X^{(r)}$ in $\A$ centred observables,  
 $$
   \lim_{N\to\infty} \phi_\infty\Bigl(F_N(X^{(1)}) 
   F_N(X^{(2)}) \cdots F_N(X^{(r)})\Bigr)=\left\{ 
   \eqalign{
   &0 \hskip 5truecm  r=2n+1 \cr
   &{1\over n!} {\sum_\nu}^{(2)} \phi_\infty\Bigl(X^{(1)}_{\nu(1)} 
   \cdots X^{(2n)}_{\nu(2n)}\Bigr) \;\; r=2n.
  }\right.
 $$
 $\sum_\nu^{(2)}$ means that we have to sum over all partitions $\nu$
 of $\{1,2, \ldots, 2n\}$ into pairs $\bigl((\alpha_1,\beta_1)$,
 $(\alpha_2,\beta_2)$, $\ldots, (\alpha_n,\beta_n) \bigr)$ i.e.\ we
 choose sites  $\alpha_j< \beta_j$ such that $\nu(\alpha_j)=
 \nu(\beta_j)= j$ with $j$ running from 1 to $n$.  
}

\noindent
{\bf Proof:}  
 We have to compute the limit for large $N$ of
 $$
   \phi_\infty\Bigl(F_N(X^{(1)}) F_N(X^{(2)}) \cdots F_N(X^{(r)})\Bigr)=
   {1\over N^{r/2}} \sum_{{\bf k}\in\{1,\ldots,N\}^r}
   \phi_\infty(X^{(1)}_{k_1} \cdots  X^{(r)}_{k_r})\ ,
 \eqno(21)  
 $$
 where ${\bf k}=\{k_1, \ldots, k_r\}$. We first concentrate on the
 contributions to~(21) where at least one of the indices $k_1, k_2,
 \ldots, k_r$ appears only once and show that all of them vanish. Let
 $k_p$ be such an index.  Due to strong compatibility, we may always
 assume that the map $i\mapsto k_i$ transform $\{1,2,\ldots,r\}$ into
 $\{1,2,\ldots,s\}$, with $s\le r$ due to possible repetitions of an
 index $k_i$. However, by hypothesis, the index $k_p$ appears  just
 once, meaning that the time $t_{k_p}$ does not  explicitly appear in 
 the words
 $$
   X^{(1)}(t_{k_1}) X^{(2)}(t_{k_2})\cdots X^{(p-1)}(t_{k_{p-1}})
   \qquad\hbox{and}\qquad 
   X^{(p+1)}(t_{k_{p+1}})X^{(p+2)}(t_{k_{p+2}})\cdots X^{(r)}(t_{k_r})\ .
 $$    
 Let $X^{(p)}$ be a centred observable and use that $\phi_\infty$ is
 defined by~(14) and that the dynamical system $(\A,\Theta,\phi)$ 
 satisfies condition~(20). Then, Lemma~1 guarantees that, given any
 $\epsilon>0$, we can find a corridor function
 $\Delta_d^s(t_1,t_2,\ldots,t_s)$ with $d$ so large that
 $$\eqalign{
   &\Bigl|\phi_\infty\Bigl(X^{(1)}_{k_1} \cdots X^{(p-1)}_{k_{p-1}} 
   X^{(p)}_{k_p} X^{(p+1)}_{k_{p+1}}\cdots X^{(r)}_{k_r}\Bigr)\Bigl|=\cr 
   &\Bigr|\avg\Bigl(t_s\mapsto \cdots\avg\Bigl(t_2\mapsto 
   \avg\Bigl(t_1\mapsto\cr 
   &\qquad\phi\Bigl(X^{(1)}(t_{k_1})\cdots
   X^{(p-1)}(t_{k_{p-1}})X^{(p)}(t_{k_p})
   X^{(p+1)}(t_{k_{p+1}})\cdots X^{(r)}(t_{k_r})\Bigr)\Bigr)\Bigr)\cdots\Bigr)
   \Bigr|\le\epsilon\ . 
 }$$ 
 Therefore, the only non-zero contributions come from terms where none
 of the subindices $k_j$ appears alone. \nl
 Next, we prove that, in the limit of large $N$, all those
 contributions vanish which come from terms where all indices appear
 twice and at least one thrice. As $\phi_\infty$ is a state, we have
 the a priori estimate   
 $$
   |\phi_\infty\Bigl(X^{(1)}_{k_1} \cdots
   X^{(r)}_{k_r}\Bigr)| \le \prod_{\ell=1}^r\|X^{(\ell)}\|\ . 
 $$ 
 Let us partition $\{1,2,\ldots, r\}$ into $s$ groups of equal indices 
 and let $r_1, r_2, \ldots, r_s$ denote the number of elements in each
 such group.  By assumption, $r_j\ge 2$, at least one of the $r_j\ge 3$
 and $r_1+ r_2+ \cdots+ r_s= r> 2s$. The contribution of all these
 terms can be estimated as follows: any partition ${\bf
 k}: \{1,2,\ldots,r\}\mapsto\{k_1,k_2,\ldots,k_s\}$, where
 $k_j\in\{1,2,\ldots,N\}$, can be written as ${\bf k}=\theta\circ\nu$,
 with $\nu:\{1,2,\ldots,r\}\mapsto\{1,2,\ldots,s\}$ a partition and
 $\theta:\{1,2,\ldots, s\}\mapsto\{1,2,\ldots, N\}$ an order preserving
 injection. There are ${N\choose s}$ of such order preserving
 injections. Thus, the contribution of the type considered is bounded
 from above by   
 $$
   {1\over N^{r/2}}\, \sum_{s<{r\over2}} A_s\, {N\choose s}\, 
   \prod_{\ell=1}^r\|X^{(\ell)}\|\ .
 $$
 $A_s$ is the number of partitions $\nu: \{1,2,\ldots, r\}\mapsto
 \{1,2,\ldots, s\}$ such that each $1<j<s$ appears at least twice. Such
 an upper bound tends to zero when $N$ tends to infinity. \nl
 Therefore, we remain with contributions of the form 
 $\phi_\infty\Bigl(X^{(1)}_{k_1} \cdots X^{(r)}_{k_r}\Bigr)$ 
 where the $k_j$ run
 from $1$ to $N$ and each of them appears exactly twice, whence $r$ has
 to be even, say $r=2n$.  These contributions are given by 
 $\phi_\infty\Bigl(X^{(1)}_{\nu(1)} \cdots X^{(2n)}_{\nu(2n)}\Bigr)$, 
 where $\nu:\{1,\ldots,2n\}\mapsto\{1,\ldots,n\}$ is a (not necessarily
 ordered) pair partition, that is, for any $j\in\{1,2,\ldots,n\}$,
 there exist two, and only two, indices $(\alpha_j, \beta_j)\in
 \{1,2,\ldots,2n\}$ such that $\nu(\alpha_j)=\nu(\beta_j)=j$. 
 In fact, as before: any pair partition ${\bf k}: \{1,2, \ldots,
 2n\}\mapsto \{k_1, k_2, \ldots, k_n\}$, where $k_j\in \{1,2, \ldots,
 N\}$, can be written as ${\bf k}= \theta\circ \nu$ with $\nu: \{1,2,
 \ldots, 2n\}\mapsto \{1,2, \ldots, n\}$ a pair partition and $\theta:
 \{1,2, \ldots, n\}\mapsto \{1,2, \ldots, N\}$ an order preserving
 injection. There being $N\choose n$ of such injections,
 the sum of contributions from generic pair partitions ${\bf
 k}:\{1,2, \ldots, 2n\}\mapsto\{1,2, \ldots, N\}$ can be simplified to
 $$
   \sum_{{\bf k}}\phi_\infty\Bigl(X^{(1)}_{k_1}  X^{(2)}_{k_2}\cdots 
   X^{(r)}_{k_r}\Bigr) ={N\choose n}\ {\sum_\nu}^{(2)}
   \phi_\infty\Bigl(X^{(1)}_{\nu(1)} X^{(2)}_{\nu(2)}\cdots
   X^{(2n)}_{\nu(n)}\Bigr)\ ,
 $$
 where $\nu$ is any pair partition of $\{1,2,\ldots, 2n\}\mapsto
 \{1,2,\ldots,n\}$.
 As $\lim_N N^{-n}{N\choose n}=1/n!$, the result follows.
\QED

\noindent
{\bf Remarks}
\item{a.}
 Condition~(20) is only sufficient to obtain the {\ssf central
 limit theorem} of Proposition~3.
 The result also follows from weak clustering if the asymptotic state 
 $\phi_\infty$ is permutation invariant. Indeed, we could then 
 average first over the time $t_{k_p}$ that appears only once
 and use weak clustering to conclude that this  average is zero.
 According to Proposition~2, this occurs when  $(\A,\Theta,\phi)$ is strongly
 clustering, hence weakly clustering and $\phi_\infty$ automatically permutation
 invariant.
\item{b.}
 Condition~(20) is implied by strong clustering~(6.a) because of the
 equivalence of~(6.a) and~(6.b) and implies weak clustering~(5). In fact, with
 $\widetilde{Y}:=Y-\phi(Y)$, condition~(20) means that
 $$\lim_{t\to\infty}\phi\Bigr(X\widetilde{Y}(t)Z\Bigl)=
   \lim_{t\to\infty}\phi\Bigr(X(-t)\widetilde{Y}Z(-t)\Bigl)=0\ .
 $$

\medskip
\noindent
As far as fluctuations are concerned, permutation invariance~(16) implies

\noindent
{\bf Corollary~1.}
{\it 
 Let the dynamical system $(\A,\Theta,\phi)$ be weakly clustering. Let the
 multiple average~(14) exist and define an asymptotic state $\phi_\infty$ 
 which is permutation invariant in the sense of~(16).
 Then, with $X^{(1)}, \ldots,
 X^{(r)}$ centred observables of $\A$, 
 $$
   \lim_{N\to\infty} \phi_\infty\Bigl(F_N(X^{(1)}) 
   F_N(X^{(2)}) \cdots F_N(X^{(r)})\Bigr)=\left\{ 
   \eqalign{
   &0 \hskip 5truecm  r=2n+1 \cr
   &{\sum_{\nu,\, {\rm ord}}}^{(2)} \phi_\infty\Bigl(X^{(1)}_{\nu(1)} 
   \cdots X^{(2n)}_{\nu(2n)}\Bigr) \;\; r=2n\ ,
   }\right.
 $$
 where ${\sum_{\nu,\, {\rm ord}}}^{(2)}$ means that the sum is   over all
 {\ssf ordered} pair partitions $\nu= \bigl( (\alpha_1,\beta_1)$,
 $(\alpha_2, \beta_2)$, $\ldots, (\alpha_n, \beta_n)\bigr)$
 of  $\{1,2, \ldots, 2n\}$, i.e.\ we choose sites 
 $\alpha_j< \beta_j$ such that $\nu(\alpha_j)= \nu(\beta_j)= j$ with $j$ 
running
 from 1 to $n$ and $\alpha_1< \alpha_2< \cdots< \alpha_n$.  
}

\noindent
{\bf Proof:}
 An easy consequence of Proposition~3.
\QED 

\noindent
{\bf Corollary~2.}
{\it 
 If the dynamical system $(\A,\Theta,\phi)$ is strongly clustering, then
 the fluctuations $F_N(X)$ of observables $X= X^*\in\A$ such that
 $\phi(X)=0$ and $\phi(X^2)=\sigma^2$
 tend to Gaussian random variables with zero mean and variance
 $\sigma$.
}

\noindent
{\bf Proof:}
 Since $\phi(X)=0$ implies $\phi_\infty(X_j)=0$, for all $j\in\Nl_0$,
 we are in fact dealing with centred words in $\A_\infty$.
 Because of Proposition~2, we can apply Corollary~1, using the notation
 for pair partitions introduced there,
 to compute the {\ssf even moments} (the odd ones are zero) 
 $$
   M_{2n}:=\lim_N\phi_\infty\Bigl(F_N(X)^{2n}\Bigr)\ .
 $$
 As in Proposition~2, it follows that
 $$
   \phi_\infty\Bigl(X_{\nu(1)} X_{\nu(2)}\cdots X_{\nu(2n)}\Bigr)=
   \prod_{j=1}^n\phi_\infty\Bigl(X_{\nu(\alpha_j)}X_{\nu(\beta_j)}\Bigr)=
   \sigma^{2n}\ ,
 $$
 where we have used that 
 $\phi_\infty\Bigl(X_{\nu(\alpha_j)}X_{\nu(\beta_j)}\Bigr)=
 \phi_\infty\Bigl((X^2)_j\Bigr)=\phi(X^2)=\sigma^2$.
 Since the number of pair partitions 
 $\nu:\{1,2,\ldots,2n\}\mapsto\{1,2,\ldots,n\}$ is 
 $\prod_{j=0}^n{2(n-j)\choose 2}=(2n)!/2^n$, we get
 $$
   M_{2n}=(2n-1)!!\ \sigma^{2n}\ ,
 $$
 that is the $2n$-th moment of the Gaussian distribution $g_{0,\sigma}(t)$ 
 with zero mean and variance $\sigma$: 
 $g_{0,\sigma}(t)=1/\sqrt{2\pi\sigma^2}\exp(-t^2/2\sigma^2)$.
\QED 

Proposition~3 is sufficient to guarantee that the linear functional
$\Phi$ defined as a pointwise limit by~(19) is positive  on the free
$\ast$-algebra $\F$ generated by all possible fluctuations $F(X)$, 
$X\in\A$, where $F(X)^*:=F(X^*)$. Furthermore, one can look for
concrete Hilbert space representations  of the abstract couples
$(\F,\Phi)$. More precisely, one may search for a structure determined
by creation and annihilation operators $a^*$ and $a$ satisfying
certain  commutation relations~[14] and for a ``ground'' state
$\Omega$ such that 
$$
  \langle \Omega, (a+a^*)^n \Omega\rangle= \Phi(F(X)^n)\ .
$$ 
It is, however, clear that associating definite probability
distributions and representations to fluctuations very much depends  on
how well-behaved the even moments in Proposition~3 are.  There are a
few cases, Corollary~2 being one of them,  where a definite structure
emerges. We present two of them, but first recall the notion of {\ssf
non-crossing pair partitions}.

Let $\nu$ be a pair partition of $\{1,2,\ldots,2n\}$ into $n$ pairs 
$(\alpha_j,\beta_j)$, with $\alpha_j<\beta_j$, such that 
$\nu(\alpha_j)=\nu(\beta_j)=j$, $j\in\{1,2,\ldots,n\}$.
A {\ssf crossing} occurs when 
$$
  \alpha_j<\alpha_k<\beta_j<\beta_k\quad\hbox{for some}\quad 
   j,k\in\{1,2,\ldots,n\}\ .
$$
{\lessblank
Denoting by $c(\nu)$ the number of crossings in a pair partition $\nu$,
$\nu$ is non-crossing if $c(\nu)=0$. If $\nu$ is a non-crossing pair
partition of $\{1,2,\ldots,2n\}$, then its pairs $(\alpha_j,\beta_j)$,
$j=1,2,\ldots,n$, are nested, that is if $\alpha_j<\alpha_k<\beta_j$
for some $j,k$, then also $\beta_k<\beta_j$.
\item{$\bullet$}
Bosonic Brownian motion
$$
  \phi_\infty\Bigl(X^{(1)}_{\nu(1)} \cdots X^{(2n)}_{\nu(2n)}\Bigr)=
  \prod_{k=1}^n\phi\Bigl(X^{(\alpha_k)}X^{(\beta_k)}\Bigr)\ .
$$
can be represented in terms of Bosonic creation and annihilation
operators, i.e.\ $aa^*-a^*a=\idty$. 
\item{$\bullet$}
Free Brownian motion
$$
 \phi_\infty\Bigl(X^{(1)}_{\nu(1)} \cdots X^{(2n)}_{\nu(2n)}\Bigr)=
 \delta_{0,c(\nu)} \prod_{k=1}^n \phi\Bigl(X^{(\alpha_k)}
 X^{(\beta_k)}\Bigr)\ .
$$
can be represented on the full Fock space in terms of creation and
annihilation operators satisfying $aa^*=\idty$.

}
\medskip
In the case of the Bosonic Brownian motion, the
ground state distribution is Gaussian.  In the case of the Free
Brownian motion the ground state distribution is the semicircle law and
to it only non-crossing pair partitions contribute~[7]. Indeed, the
semicircle distribution
$$
  \gamma_{0,1}(t)={1\over 2\pi}\sqrt{4-t^2}\ ,\qquad |t|\le 2
$$
has vanishing odd moments, whereas the even ones are given by the
Catalan numbers $C_n$: $M_{2n}={2n\choose n}/(n+1)= C_n$. The $C_n$
are defined by the recursion relation
$$
  C_0=1\ ,\qquad C_n=\sum_{k=1}^n\ C_{k-1}\ C_{n-k}\ .
\eqno(22)
$$ 
The Catalan numbers count in how many different ways one can partition
$\{1,2,\ldots,2n\}$ in ordered non-crossing pairs  $\{(\alpha_1,
\beta_1), (\alpha_2, \beta_2), \ldots, (\alpha_n, \beta_n)\}$, with
$\alpha_1< \alpha_2< \cdots< \alpha_n$.

In the following proposition, we exhibit conditions on the dynamical
system $(\A,\Theta,\phi)$ and on the procedure of averaging multi-time
correlation functions such that a free statistics for time-asymptotic 
fluctuations arises. We start by observing that the explicit rule~(14)
for computing averages  has a natural product structure: if  ${\bf
t}\mapsto f_1(t_{i_1},t_{i_2}, \ldots, t_{i_m})$ and   ${\bf t}\mapsto
f_2(t_{j_1}, t_{j_2},\ldots, t_{j_n})$  are uniform limits of
multi-time correlation functions, then
$$
  \{i_1, i_2, \ldots, i_m\} \cap \{j_1, j_2, \ldots, j_n\}= \emptyset\
   \Longrightarrow \ \avg(f_1 f_2)= \avg(f_1)\, \avg(f_2)\ .
\eqno(23)
$$   
Notice, however, that an average is generally not specified if we impose the
product property and specify all single-time averages $\avg$. Indeed,
e.g.\ a bounded function $f$ over $\Ir^2$ is not necessarily the
uniform limit of linear combinations of products of bounded functions
${t_1,t_2}\mapsto h(t_1) g(t_2)$ over $\Ir$.

\noindent
{\bf Proposition~4.}
{\it 
 Let $(\A,\Theta,\phi)$ satisfy condition~(20) and let the multiple average~(14)
 exist and define an asymptotic state $\phi_\infty$ on $\A_\infty$.
 Let us also assume that for any time independent choice of
 observables $A$ and $C$  and centred observables $X$, $Y$ and $B$ 
 $$
   \avg\Bigl(t\mapsto \phi\Bigl(A X(t) B Y(t) C\Bigr)\Bigr)= 0\ .
 \eqno(24)
 $$ 
 Then, if  $X^{(1)}, \ldots, X^{(2n)}$ are centred observables in $\A$,  
 $$
   \lim_{N\to\infty} \phi_\infty\Bigl(F_N(X^{(1)}) F_N(X^{(2)}) \cdots
   F_N(X^{(2n)}) \Bigr)= {1\over n!}{\sum_{\nu}}^{(2)} 
   \delta_{0,c(\nu)} \prod_{k=1}^n \phi\Bigl( X^{(\alpha_k)}
   X^{(\beta_k)}\Bigr)\ , 
 $$ 
 where the sum extends over all pair partitions $\nu= \bigl((\alpha_1,
 \beta_1), (\alpha_2, \beta_2), \ldots, (\alpha_n,\beta_n)\bigr)$ of
 $\{1,2,\ldots,2n\}$. If the asymptotic state $\phi_\infty$ is also
 permutation invariant, then
 $$
   \lim_{N\to\infty} \phi_\infty\Bigl(F_N(X^{(1)}) F_N(X^{(2)}) \cdots
   F_N(X^{(2n)}) \Bigr)= {\sum_{\nu,\, {\rm ord}}}^{(2)} 
   \delta_{0,c(\nu)} \prod_{k=1}^n \phi\Bigl( X^{(\alpha_k)}
   X^{(\beta_k)}\Bigr)\ , 
 $$
 where the sum now extends over all ordered pair partitions $\nu$ of
 $\{1,2,\ldots,2n\}$.
}

\noindent
{\bf Proof:} 
As the asymptotic state $\phi_\infty$ is defined according to~(14), we
know it to be strongly compatible and endowed with the product
structure~(23). It need not be, however, permutation invariant. In any
case, on the basis of Proposition~3, we only have to consider even
moments of order $2n$. More precisely, we can restrict to expectations
of the form  $\phi_\infty\Bigl(X^{(1)}_{\nu(1)}\cdots
X^{(2n)}_{\nu(2n)}\Bigr)$, where $\nu$ is a pair partition of
$\{1,2,\ldots, 2n\}$. In order to compute such an expectation, we must
perform successive time-averages as in~(14). We begin with averaging
with respect to $t_1$
$$
  \avg\Bigl(t_1\mapsto \phi\Bigl(X^{(1)}(t_{\nu(1)}) X^{(2)}(t_{\nu(2)})\cdots 
  X^{(2n)}(t_{\nu(2n)}) \Bigr) \Bigr)\ .
$$ 
Either the time $t_1$ appears in two consecutive words $X^{(j)}$ and
$X^{(j+1)}$, in which case we can use weak clustering to obtain
$$\eqalign{
  &\avg\Bigl(t_1\mapsto \phi\Bigl(X^{(1)}(t_{\nu(1)}) X^{(2)}(t_{\nu(2)})\cdots 
  X^{(2n)}(t_{\nu(2n)}) \Bigr) \Bigr) \cr
  &\quad = \phi(X^{(j)} X^{(j+1)})\, \phi\Bigl(
  X^{(1)}(t_{\nu(1)})\cdots X^{(j-1)}(t_{\nu(j-1)})X^{(j+2)}(t_{\nu(j+2)})\cdots
  X^{(2n)}(t_{\nu(2n)}) \Bigr) \Bigr)\ .
}$$  
Or, the two words at time $t_1$ are separated by one or more other
words at different times: $\phi\Bigl(w\, X^{(\alpha_1)}(t_1)\, w'\,
X^{(\beta_1)}(t_1)\, w''\Bigr)$. In this case we use condition~(24) as
follows. Notice that $w'$ is not necessarily centred and so we write 
$$
  w'= (w'-\phi(w')\idty)+ \phi(w')\idty \ .
$$
Then, remembering that the time $t_1$ does not appear in $w,\, w',\,
w''$, condition~(24) and weak clustering imply 
$$\eqalignno{
  &\avg\Bigl(t_1\mapsto \phi\Bigl(w\, X^{(\alpha_1)}(t_1)\, w'\, 
  X^{(\beta_1)}(t_1)\, w''\Bigr)\Bigr)\cr
  &\quad= \phi(w')\, \avg\Bigl(t_1\mapsto 
  \phi\Bigl(w \Bigl(X^{(\alpha_1)} X^{(\beta_1)}\Bigr)(t_1)\, w''\Bigr)\Bigr)\cr
  &\quad=\phi(X^{(\alpha_1)}X^{(\beta_1)})\, \phi(w')\, \phi(w w'')\ .
  &(25)
}$$
Consider now the average with respect to $t_2$. This is either a
similar average as that with respect to $t_1$ or else the time $t_2$
appears separately in each of the factors $\phi(w')$ and $\phi(w
w'')$ in~(25). If so, the average with respect to $t_2$ vanishes
because of weak clustering and the centredness of the $X$'s. Repeating
this argument for all the time averages, we see that only nested pairs
i.e.\ non-crossing pair partitions, contribute whence the result.
\QED 

Proposition~4 and permutation invariance of the asymptotic state
provide us with sufficient conditions on the correlation
functions to yield fluctuations which are free random variables and
hence semicircularly distributed.

\noindent
{\bf Corollary~3.}
{\it
 Let the dynamical system $(\A,\Theta,\phi)$ satisfy conditions~(20) 
 and~(24) and the multiple average~(14) exist and define a
 permutation invariant asymptotic state $\phi_\infty$ as in~(16).
 Then, the fluctuations $F_N(X)$ 
 of observables
 $X= X^*\in \A$ such that $\phi(X)=0$ and $\phi(X^2)=\sigma^2$ tend to
 semicircularly distributed random variables with zero mean and
 variance $\sigma$.
}

\noindent
{\bf Proof:}
We can apply the previous proposition to compute 
the even moments (the odd ones are zero) 
$$
  M_{2n}:=\lim_N\phi_\infty\Bigl(F_N(X)^{2n}\Bigr)=\ C_n\sigma^{2n}\ ,
$$
where $C_n$ are the Catalan numbers~(22). On the other hand, the
latter moments are the non-zero moments of the semicircle distribution 
$\gamma_{0,\sigma}(t)=\sqrt{4\sigma^2-t^2}/2\pi\sigma^2$ on
$|t|\le2\sigma$.
\QED 

\noindent
{\bf Remarks}
\item{a.}
Referring to the types of clustering we presented in Section~2, clustering in
the mean~(4) would not be sufficient to yield the result of Proposition~4, 
while strong clustering~(6.a) is incompatible with condition~(24)
and with freeness, as already observed in Section~3. 
In the proof of Proposition~4, weak clustering~(5) has been used in an
essential way. Indeed, suppose that we assume instead of weak
clustering only clustering in the mean. In the process of centring the
observable $w'$, sandwiched  between two equal time observables
$X^{(i_k)}(t_k)$ and $X^{(j_k)}(t_k)$, some other time $t_\ell$ might
appear once in both $w'$ and in $ww''$. Therefore, we should consider 
an average of a product $t\mapsto f(t)g(t)$ which is generally unequal 
to the product of the averages
$$
  \avg\Bigl( t\mapsto f(t)g(t)\Bigr)\ne \avg\Bigl( t\mapsto f(t)\Bigr)
  \avg\Bigl( t\mapsto g(t)\Bigr) \ .
$$
\item{b.}
Conditions~(20) and~(24) are independent.
Obviously, the latter does not imply the first one.
As a counterexample to $(20)\Rightarrow(24)$, consider the algebra $\A$
generated by linear combinations of
the unitary operators $W(\p)$, $\p=(p_1,p_2)\in\Ir^2$, such that
$$W(\p)^*=W(-\p)\,,\quad
  W(\p)W(\q)={\rm e}^{i\pi\theta\sigma(\p,\q)}W(\p+\q)\ ,
$$
where $\theta\in[0,1)$ and $\sigma(\p,\q)=p_1q_2-p_2q_1$.
By equipping $\A$ with the tracial state $\phi(W(\p))=\delta_{\p,\o}$ 
and the automorphisms $\Theta^t(W(\p)=W(T^t\p)$, where $T$ is the typical
$2\times 2$ matrix of the hyperbolic dynamics on the two-dimensional torus,
one obtains the class of quantized cat-maps studied in~[15].
A thorough examination of such models in the framework presently developed 
will be the subject of a forthcoming paper~[16].
Here, it is sufficient to mention that there are values of $\theta$ for which
$\lim_t\theta\sigma(T^t\p,\q)\mod{1}=\beta\Delta(\p,\q)$ with 
$\beta$ and $\Delta(\p,\q)$ two not necessarily zero quantities. \nl
One calculates
$\phi\Bigr(W(T^{t_1}\p)W(T^{t_2}\q)W(-\p)W(-T^{t_2}\q)\Bigl)=
  {\rm e}^{2\pi i\theta\sigma(T^{t_2}\q,\p)}\delta_{T^{t_1}\p,\p}$. 
Thus,
$\avg\Bigr(t\mapsto\phi\Bigr(W(\p)W(T^t\q)W(-\p)
  W(-T^t\q)\Bigl)\Bigl)={\rm e}^{2\pi i\beta\Delta(\q,\p)}$,  
whereas\nl
$\lim_{t_1\,,t_2\to\infty}\phi\Bigr(W(T^{t_1}\p)
W(T^{t_2}\q)W(-\p)W(-T^{t_2}\q)\Bigl)=0$.

\beginsection{5 Examples}

We implement in this section the constructions of above for a few simple
model systems. More complex dynamical systems, with richer behaviour of
correlation functions, will be considered in a forthcoming paper~[16].
As Proposition~2 and Corollary~1 deal with the asymptotic state and the
fluctuations of strongly clustering dynamical systems, we shall only 
consider here more extremely non-commuting dynamical systems.  

As a first model, we consider a probability space $(X,\mu)$ equipped
with an automorphism $\theta$ i.e.\ a measurable transformation $\theta$
of $X$, with measurable inverse $\theta^{-1}$, such that $\mu= \mu\circ
\theta= \mu\circ \theta^{-1}$. We will assume that $\theta$ is {\ssf
mixing}
$$ 
  \lim_{n\to\infty} \mu(f\, g\circ\theta^n)= \mu(f)\, \mu(g)
  \qquad\hbox{for all}\qquad
  f,\, g\in\L^1(X,\mu) \ .
$$
In the {\ssf Koopman} description of the classical dynamical system
$(X,\mu,\theta)$, one considers the Hilbert space $\H:= \L^2(X,\mu)$ and the
unitary single step evolution $Uf:= f\circ\theta$ of the ``wave
functions'' $f$. The expectation $\mu(f)$ of an $\L^\infty$ function on
$X$ is recovered as 
$$
  \mu(f)= \langle {\bf 1}, M_f\, {\bf 1} \rangle \ .
$$
In this formula $\bf 1$ denotes the constant function 1 on $X$ and
$M_f$ the multiplication operator on $\H$ by the function $f$. We now
consider a ``non-canonical'' quantization whereby the quantum evolution
of ``wave-functions'' is exactly the classical one~[17]. In such a
description, one may choose for $\A$ the algebra of compact operators
$\Cx\idty+ \K(\H)$ to which a unit has been added. The dynamics is
determined by the usual Heisenberg picture: $\Theta(X):= U\, X\, U^*$
and the reference state $\phi$ is the vector state defined by $\bf 1$.
Using the assumed mixing property of $(X,\mu,\theta)$, one readily
verifies that whenever $X\in\K(\H)$ and $w$ and $w'$ are products in
$\A_\infty$ of copies of compact operators, such that both the
rightmost letter of $w$ and the leftmost of $w'$ are different from
$j$, then
$$
  \phi_\infty(w X_j w')= \phi(X)\ \phi_\infty(w)\ \phi_\infty(w') \ ,
\eqno(26)  
$$  
even when $w$ or $w'$ contain letters pertaining to the index $j$. As
a consequence, 
$$
  \phi_\infty \Bigl( X^{(1)}_{\nu(1)} X^{(2)}_{\nu(2)} \cdots
  X^{(n)}_{\nu(n)}\Bigr)= \prod_{j=1}^n \phi(X^{(j)})
$$
whenever $X^{(j)}\in \K(\H)$ and $\phi_\infty$ is permutation
invariant. For a compact $X\in\K(\H)$ we will denote by $\tilde X$ the
element $X- \phi(X)\idty$ obtained by centring $X$. In spite of good
ergodic properties of $(X,\theta,\mu)$, its ``quantization''
$(\A,\Theta,\phi)$ is not weakly clustering as can be seen from
$$
  \lim_{t\to\infty} \phi(X \Theta^t(Y) X)= \phi(X)^2\ \phi(Y)\ne
  \phi(X^2)\ \phi(Y) \ ,
$$
$X$ and $Y$ compact. Considering fluctuations is not meaningful in such
a case. In fact, expectations of third order moments of fluctuations
$F_N(X)$ diverge with $N$ as can be seen from
$$
  \phi_\infty(F_N(X))= {1\over \sqrt N} \phi(\tilde X^3)- {N-1\over
  \sqrt N} \phi(X)\, \Bigl(\phi(X^2)- \phi(X)^2 \Bigr) \ ,
$$
with $\tilde X:= X- \phi(X)\idty$.    

As a second example, we consider a Fermionic dynamical system whose
observables are given by a CAR-algebra $\A(\H)$ over a single-particle
space $\H$. $\A(\H)$ is generated by an identity $\idty$ and by creation
and annihilation fields $\{a^*(\varphi)\mid \varphi\in\H\}$ and
$\{a(\varphi)\mid \varphi\in\H\}$ subject to the relations:
$$\eqalign{
  &\varphi\mapsto a^*(\varphi) \hbox{ is $\Cx$-linear and} \cr
  &\{a(\varphi), a(\psi)\}=0 
  \qquad\hbox{and}\qquad
  \{a(\varphi), a^*(\psi)\}= \langle \varphi,\psi \rangle \ .  
}$$
The {\ssf parity} automorphism $\pi$ on $\A(\H)$ is determined by
$\pi(a^*(\varphi)):= - a^*(\varphi)$ and a $\ast$-automorphism $\Theta$
is said to be {\ssf even} if $\Theta\circ\pi= \pi\circ\Theta$. We assume
furthermore that the dynamics determined by such an even $\Theta$ is
asymptotically ``Abelian'' in the sense that
$$
  \lim_{n\to\infty} \Bigl\|\Bigl\{a^*(\varphi),
  \Theta\bigl((a^\#(\psi)\bigr)\Bigr\}\Bigr\|= 0 \ ,
$$
where $a^\#$ denotes either $a$ or $a^*$. It is well-known that a
reference state $\phi$, invariant under an even, asymptotically Abelian
$\Theta$, is automatically even i.e.\ $\phi\circ\pi= \phi$. Finally, we
shall assume that the dynamical system is multi-clustering 
$$
  \lim_{t_j- t_{j+1}\to\infty} \phi\Bigl( X^{(1)}(t_1) X^{(2)}(t_2)
  \cdots X^{(k)}(t_k)\Bigr)= \prod_{j=1}^k \phi(X^{(j)})
$$
with $t_1> t_2> \cdots> t_k$. The asymptotics of multi-correlation
functions can be computed along the same lines as that for the strongly
clustering case and we obtain when each of the $X^{(j)}$ is either even
or odd
$$
  \phi_\infty \Bigl(X^{(1)}_{\nu(1)}  X^{(2)}_{\nu(2)} \cdots
   X^{(n)}_{\nu(n)}\Bigr)= \epsilon \prod_{p=1}^k \phi_\infty\Bigl( 
   \overrightarrow{\prod}_{i\in I_p} X^{(i)}\Big) \ .
\eqno(27)   
$$
$I_p$ is, as in Proposition~2, the set of all indices $j$ such that
$\nu(j)=p$ and $\epsilon$ is either 1 or $-1$ according to whether    
an even or odd permutation is needed to permute the odd $X^{(j)}$ 
appearing in $X^{(1)}_{\nu(1)} X^{(2)}_{\nu(2)} \cdots X^{(n)}_{\nu(n)}$
into the order in which they appear in the product at the right-hand
side of~(27). This amounts to saying that $\phi_\infty$ is the Chevalley
product of a countable number of copies of $\phi$ on the twisted tensor
product $\A(\oplus^{\Nl_0} \H)$ of $\Nl_0$ copies of $\A(\H)$ with itself.  
The fluctuations of creation and annihilation fields are now
straightforwardly computed, yielding
$$
  \lim_{N\to\infty} \phi_\infty\Bigl( F_N(a^\#(\varphi_1))
  F_N(a^\#(\varphi_2)) \cdots F_N(a^\#(\varphi_n))\Bigr)= \phi_{\rm
  QF}\Bigl(a^\#(\varphi_1) a^\#(\varphi_2) \cdots a^\#(\varphi_n)\Bigr)
  \ .
$$  
$\phi_{\rm QF}$ is the ``quasi-free projection'' of the state $\phi$,
namely the quasi-free state on $\A(\H)$ determined by the covariance 
$$
  (\varphi,\psi)\mapsto \phi(a^\#(\varphi)a^\#(\psi)) \ .
$$
We recover hereby the Fermionic central limit theorem of~[18].  

In its most basic form, the ``free shift'' is a quantum shift
$$
  \Theta(e_i):= e_{i+1}, \qquad i\in\Ir
$$
on a set of generators $\{e_i\mid i\in\Ir\}$ that satisfy the relations
$$
  e_i^*= e_i
  \qquad\hbox{and}\qquad 
  e_i^2=\idty, \qquad i\in\Ir \ .
$$
The algebra $\A$ of observables is the universal C*-algebra
generated by $\idty$ and the $e_i$. The finite linear combinations of
monomials of the type $e_{i_1}e_{i_2}\cdots e_{i_m}$ with $i_k\ne
i_{k+1}$ form a dense $\ast$-subalgebra $\A_0$ of $\A$. The product of two
monomials determined by ordered index sets $\{i_1, i_2, \ldots, i_m\}$
and $\{j_1, j_2, \ldots, j_n\}$ is the monomial corresponding to the
index set obtained by first concatenating $\{i_1, i_2, \ldots, i_m\}$
and $\{j_1, j_2, \ldots, j_n\}$ and then omitting those indices that
appear twice in subsequent positions. The identity corresponds to the
monomial with empty index set and the adjoint of a monomial is the
monomial with reversed index set. If there are no
preferred observables to single out apart from the identity, a
meaningful reference state is the state $\phi$ on $\A$ 
$$
  \phi(e_{i_1}e_{i_2}\cdots e_{i_m})= 0 \hbox{ if } m>0
  \qquad\hbox{and}\qquad
  \phi(\idty)=1 \ .
\eqno(28)  
$$ 
The dynamical system $(\A,\Theta,\phi)$ is weakly, but not strongly,
clustering. It is quite straightforward to compute the asymptotic state
$\phi_\infty$: it is the free product $\ast_{i\in\Nl_0} \phi$ of 
copies of $\phi$. As each centred element in $\A_0$ is a finite linear
combination of monomials with non-trivial dependence set, it suffices
to show that for $n=1,2, \ldots$ the expectation of an element 
$X^{(1)}_{\nu(1)} X^{(2)}_{\nu(2)} \cdots X^{(n)}_{\nu(n)}$ in the
state $\phi_\infty$ vanishes, where each $X^{(k)}$ is a non-trivial
monomial and consecutive $\nu(k)$ are different. Clearly, when all
differences $|t_k- t_\ell|$ for $k\ne \ell$ appearing in the list of
$\nu(j)$'s become sufficiently large, there is no possible
simplification in the monomial $X^{(1)}(t_{\nu(1)}) X^{(2)}(t_{\nu(2)})
\cdots X^{(n)}(t_{\nu(n)})$ due to the rule $e_i^2= \idty$ and because
of~(28) $\phi(X^{(1)}(t_{\nu(1)}) X^{(2)}(t_{\nu(2)}) \cdots
X^{(n)}(t_{\nu(n)}))= 0$ and so $\phi_\infty(X^{(1)}_{\nu(1)}
X^{(2)}_{\nu(2)} \cdots X^{(n)}_{\nu(n)})= 0$ too. The state
$\phi_\infty$ is permutation invariant and satisfies even a
strengthened Condition~(24), where the average is replaced by a limit.
Temporal fluctuations of centred self-adjoint observables are
therefore, by Proposition~4, semicircularly distributed.

\noindent
{\bf Acknowledgements:} 

F.B. acknowledges financial support from the Onderzoeksfonds K.U.Leuven
F/97/60 and the Italian I.N.F.N. and M. De Cock acknowledges financial
support from FWO-project G.0239.96.

{\parindent=0pt
\noindent
{\bf References}
\medskip

[1]
Casati~G., Chirikov~B.V., Izrailev~F.M., Ford~J.:
{\it Stochastic behaviour of a quantum pendulum under a periodic
perturbation},
Lecture Notes in Physics {\bf 93}, 334--352, New York: Springer Verlag, 1979

[2]
Haake~F., Kus~M., Scharf~R.:
Classical and quantum chaos for a kicked top.
Z.\ Phys.\ B{\bf 65}, 381--395 (1987)

[3]
Berry~M.V., Balazs~N.L., Tabor~M., Voros~E.:
Quantum maps.
Ann.\ Phys.\ {\bf 122}, 26--63 (1979) 

[4]
Balazs~N.L., Voros~A.:
The quantized baker's transformation.
Ann.\ Phys.\ {\bf 190}, 1--31 (1989)

[5]
Casati~G., Chirikov~B.V.: 
{\it Quantum Chaos}. 
Cambridge: Cambridge University Press, 1995

[6]
Benatti~F., Fannes~M.:
{\it Statistics and quantum chaos},
J.\ Phys.\ A, in press 

[7]
Speicher~R.:
Generalized statistics of macroscopic fields.
Lett.\ Math.\ Phys.\ {\bf 27}, 97--104 (1993) 

[8]
Voiculescu~D.V., Dykema~K.J., Nica~A.: 
{\it Free Random Variables}. 
Providence,RI: AMS 1992  

[9]
Emch~G.G.:
{\it Algebraic Methods in Statistical Mechanics and Quantum Field
Theory}.
New York: Wiley, 1974

[10]
Narnhofer~H., Thirring~W.:
Mixing properties of quantum systems.
J.\ Stat.\ Phys.\ {\bf 57}, 811--825 (1989)

[11]
Greenleaf~F.P.:
{\it Invariant Means on Topological Groups}.
New York: Van Nostrand Reinhold, 1969

[12]
Goderis~D., Verbeure~A., Vets~P.:
Non-commutative central limits.
Probab.\ Th.\ Rel.\ Fields {\bf 82}, 527--544 (1989)

[13]
Speicher~R., Waldenfels~W.~von:
{\it A general central limit theorem and invariance principle},
Quantum Probability and Related Topics IX, 371--387 
Singapore: World Scientific, 1994

[14]
van~Leeuwen~H., Maassen~H.:
A $q$-deformation of the Gauss distribution.
J.\ Math.\ Phys.\ {\bf 36}, 4743--4756 (1996) 

[15]
Benatti~F., Narnhofer~H., Sewell~G.L.:
A non-commutative version of the Arnold cat map.
Lett. Math. Phys. {\bf 21}, 157--192 (1991)

[16]
Andries~J., Benatti~F., De~Cock~M. and Fannes~M.:
Dynamical fluctuations in quantized toral automorphisms, in preparation

[17]
Berry~M.:
{\it True quantum chaos? An instructive example},
Proceedings of the Yukawa symposium, Tokyo (1990)

[18]
Hudson~R.L.:
A quantum mechanical central limit theorem for anti-commuting
observables.
J.\ Appl.\ Prob.\ {\bf 10}, 502--509 (1973) 
}
\bye